\documentclass[aps,prl,preprint,groupedaddress]{revtex4-1}
\usepackage{amsmath,amssymb}
\usepackage[dvips]{graphicx}
\usepackage{booktabs}
\usepackage{subfigure}

\begin{document}
\preprint{arXiv:1109.3105[hep-ph]}
\title{Resolving the LSND anomaly  by  neutrino diffraction}
\author{Kenzo Ishikawa and Yutaka Tobita}
\affiliation{Department of Physics, Faculty of Science, Hokkaido
University Sapporo 060-0810, Japan}
\date{\today}
\begin{abstract}
In the charged pion decay, a neutrino is produced in pair with  a
 charged lepton and 
 they  have the same production rate. In this paper we show that  
neutrinos have their own  space-time correlations in a wide 
area and are detected in a different manner from  charged leptons, 
owing to extremely small mass. 
The neutrino flux   reveals  a unique interference effect in the  
form of diffraction of non-stationary waves. The diffraction component 
of the flux shows  a slow 
position-dependence and  leads to an electron  neutrino at short 
base-line regions. The electron 
neutrino flux at short distances is 
attributed to the neutrino diffraction and the one at long distances is
 to the normal flavor oscillation.  The former depends upon
 the average mass-squared $\bar m^2_\nu$ and the latter depends upon the 
mass-squared difference $\delta m^2_\nu$. The LSND and the two  neutrino 
experiment~(TWN) measure $\bar m^2_\nu$  and the other experiments measure 
$\delta m^2_\nu$. Hence they are consistent with each other.
The neutrino diffraction    would supply
 valuable information on the absolute neutrino mass.   
\end{abstract}


\maketitle

\section{Neutrino diffraction and LSND.} Time evolution of a charged
 pion state is determined by  the Schr\"{o}dinger equation with 
the weak interaction Hamiltonian, and the state vector of a single 
pion  at $t=0$ becomes a superposition of one pion state and a state 
composed of a  charged lepton and a neutrino at $t=\text{T}$. A decay of
the pion is studied normally with a transition amplitude at the infinite
T of plane waves, i.e., asymptotic boundary conditions. This paper
presents that  T-dependent probabilities  of the decay products  give
 valuable informations, particularly on the neutrino, that are not found
with the asymptotic values. Since the wave function describes a time 
dependent development  of the neutrino, the wave function  is 
used to  compute 
the time-dependent probability, which   reveals  wave
phenomena. These physical quantities have been unable to observe 
experimentally with high precision before due to low statistics, but
they are becoming possible  with high statistics data of the current 
experiments.  We
study the neutrino detection
probability in a situation where the initial pion is prepared in a plane
wave using  an S-matrix of a finite time interval.       

If the  neutrino at a detector region  is highly correlated, this wave  is 
different from an isolated free wave and  is not expressed by the 
plane wave of  satisfying the asymptotic 
boundary condition of the S-matrix.  In this region,
expressing neutrinos with plane waves and studying 
their detection  processes with  the ordinary S-matrix 
\cite{LSZ,Low}, are not appropriate. An S-matrix of  a 
finite time interval defined by the M{\o}ller operator at the finite time
interval $\Omega_{\pm}(\text{T})$ as, $S[\text T]=\Omega_{-}^{\dagger}(\text T)\Omega_{+}(\text
T)$, is appropriate  to study the finite-size correction of  the neutrino
detection probability. Since wave packets decrease rapidly at  large $|{\vec
x}|$ and satisfy the asymptotic boundary conditions, the S-matrix of  a 
finite time interval is expressed  with the wave packets.  One feature of
the scattering of the finite time interval is that 
$S[\text T]$ does not commute with the free Hamiltonian $H_0$ but
satisfies 
\begin{eqnarray}
& &[S[\text T],H_0]=  i({\partial \over \partial
	\text{T}}\Omega_{-}(\text{T}))^{\dagger}\Omega_{+}(\text{T})-
	i\Omega_{-}(\text{T})^{\dagger}{\partial \over \partial \text{T}}
	\Omega_{+}(\text{T}),
\end{eqnarray}
thus the energy is not conserved by $S[\text{T}]$. Final states of having 
the different energy  from  that of the initial state  contribute to 
the finite-size correction of the transition probability. Furthermore, 
the finite-size correction has a universal property and is computed 
rigorously.
 
 We study the neutrino in the pion decay with $S[\text T]$, and find
 that  the neutrino flux has a large finite-size correction of a form 
of a diffraction. 
  We summarize our results first. The fluxes  of
the  neutrino and charged lepton   in pion decays  are defined with
 their detection rates and are given at a macroscopic distance in  the form
\begin{eqnarray}
P= P_{normal}+P_{diff}(\text T),
\label{probability}
\end{eqnarray}
where  $P_{normal}$ agrees with a normal term calculated with the standard
method  and $P_{diff}(\text T)$ is a diffraction term which is derived from
the energy-nonconserving final states.
The diffraction  components for the neutrino, $P_{diff}^{(\nu)}(\text T)$, and
for the charged lepton,  $P_{diff}^{l}$(\text T), are   expressed  by the masses
$m_{{\nu}_i},m_l$ 
and  the mixing matrix $U_{i,e}$, 
where  $i$ is  the mass
eigenstate, in the form
\begin{eqnarray}
P_{diff}^{(\nu)}(\text T)=C\text{T} \sum_i \tilde g(\text{T},\omega_{\nu_{i}})|U_{i,e}|^2,\,P_{diff}^{l}=C\text{T} \tilde g(\text{T},\omega_l),
\label{diffraction}
\end{eqnarray}
where $C$ is a constant obtained later, $\omega=\frac{m^2}{
2E}$, and $  \tilde g(\text{T},\omega)=
g(\text{T},\omega)-  \tilde g(\infty,\omega)$, where 
\begin{align}
\text{T}  g(\text{T},\omega)=-i  \int_0^{\text{T}} dt_1 dt_2  \frac{\epsilon(\delta t)}{|\delta t|}e^{i {\omega}\delta t }.\label{probability1} 
\end{align}
$\tilde g(\text{T},\omega )$ is positive definite and decreases with a
product $\text{T}\omega$. 
$\text{T}\omega_l$ becomes large and 
$\tilde g(\text{T},\omega_l )$  vanishes at a macroscopic time for the charged
leptons, but $\tilde g(\text{T},\omega_{\nu_i} )$ becomes finite for the neutrinos.
Hence $P_{diff}(\text T)$ is finite in neutrinos and depends upon an average 
mass-squared $\bar m_{\nu}^2$.  $P_{normal}$ is the normal term 
and includes flavor oscillations with the period determined by
the mass-squared differences $\delta m_{\nu}^2$.  For charged leptons 
the diffraction terms vanish and
physical quantities are computed with the normal term.

 Due to  the small  mass,
 relativistic invariance, and other features, the diffraction effect becomes
 observable in  the neutrinos. 
  The detection rates become different from their 
 production rates at  the macroscopic  distances. 
Especially because the energy-momentum is not conserved in the
 diffraction component, the helicity suppression mechanism does not work
 and  the electron  neutrino is not suppressed compared to  the muon
neutrino  in the near-detector region.
This resolves the  LSND anomaly \cite{excess-LSND} and some previous 
experiments.

\section{Position-dependent detection probability: one specie.} Now we derive  the
diffraction term \cite{Ishikawa-Tobita-prl}.  To show this effect being 
distinct from a flavor oscillation, the formula for one specie   of neutrino
is studied  first. 

We suppose that a neutrino is  observed through its incoherent 
interaction with one of nucleus. Then  its detection amplitude in the 
pion decay process   is  expressed   as, 
$T=\int d^4x \, \langle l,{\nu}
 |H_{w}(x)| \pi \rangle$. Here  a pion is prepared at a time 
 $\text{T}_\pi$,  and  a neutrino is detected at $\text{T}_\nu$, where the
 distance $c(\text{T}_\nu-\text{T}_\pi)$ is macroscopic. A  detected
 neutrino is expressed by a  wave packet that represents
the nucleon wave function  in a nucleus that neutrinos interact.  
 The neutrino wave packet  \cite {Ishikawa-Shimomura,Ishikawa-Tobita-ptp,Ishikawa-Tobita}
is  described by  the central values of the momentum and  coordinate, $(\text{T}_{\nu},
\vec{\text{X}}_{\nu})$ and the width, $\sigma_{\nu}$ 
\cite{Kayser,Giunti,Nussinov,Kiers,Stodolsky,Lipkin,Asahara}. They are
expressed in the form   
$
|\pi \rangle=   | {\vec p}_{\pi},\text{T}_{\pi}  \rangle,\ 
|l ,\nu \rangle=   |l,{\vec p}_l ;\nu,{\vec p}_{\nu},\vec{\text{X}}_{\nu},\text{T}_{\nu}          \rangle.$
The
amplitude $T$ is
written with the hadronic $V-A$ current and  Dirac spinors  in the form
\begin{align}
\label{amplitude}
T = \int d^4xd{\vec k}_{\nu}
\,N_1\langle 0 |J_{V-A}^{\mu}(x)|\pi \rangle 
\bar{u}({\vec p}_l)\gamma_{\mu} (1 - \gamma_5)\nu({\vec k}_{\nu})\nonumber\\
\times e^{ip_l\cdot x + 
ik_\nu\cdot(x - \text{X}_\nu)
 -\frac{\sigma_{\nu}}{2}({\vec k}_{\nu}-{\vec p}_{\nu})^2},  
\end{align}
where 
$N_1=ig \left({\sigma_\nu/\pi}\right)^{\frac{4}{3}}\left({m_l m_{\nu}}/{
 E_l E_{\nu}}\right)^{\frac{1}{2}}$, and  the time $t$ is
 integrated in the region $\text{T}_{\pi} \leq t$. 
${\sigma_{\nu}}$ is the size
of the neutrino wave packet and was  estimated   using  the size of a nucleus.
The Gaussian form of the wave packet is used  for the sake
of simplicity to obtain the finite-size correction in this paper.  
Its long-distance behavior is the same in general wave packets as was 
verified in \cite{Ishikawa-Tobita-prl}. 
The muon momentum of ${ p}_{\mu} \approx { p}_{\pi}-{
p}_{\nu} $ and broad tail of 
${ p}_{\mu} \neq{ p}_{\pi}-{ p}_{\nu} $
contribute to this amplitude. The former  component is the normal one
and the latter one is a diffraction component which is shown to be
computable rigorously using a light-cone singularity of relativistic 
invariant systems. 

The neutrino momentum ${\vec k}_{\nu}$ is integrated easily in Eq.\,$(\ref{amplitude})$ and the
coordinate representation of the neutrino wave is obtained, which shows
the time evolution of the neutrino wave function in the backward direction. At 
$t=\text{T}_{\nu}$, the wave function agrees with  the Gaussian function
of the center $\vec{\text{X}}_{\nu}$ and at $t
\leq \text{T}_{\nu}$ the position of the center is at ${\vec
v}_{\nu}(t-\text{T}_{\nu})+\vec{\text{X}}_{\nu}$, which overlaps with the
pion and muon
wave functions.

Integrating the space coordinates, a Gaussian function of the
momenta, which shows that the momenta are approximately conserved, are 
obtained. The time is integrated in the finite  interval $\text T=\text
T_{\nu}- \text T_{\pi}$ and the amplitude is proportional to the
T-dependent term,
\begin{eqnarray}
{\sin\left[({E(p_{\mu})+E({p_{\nu}})-E(p_{\pi}) -{\vec v}_{\nu}\cdot({\vec p}_{\nu}+{\vec
 p}_{\mu}-{\vec p}_{\pi})})\text{T}/2\right] \over E(p_{\mu})+E({p_{\nu}})-E(p_{\pi}) -{\vec v}_{\nu}\cdot({\vec p}_{\nu}+{\vec
 p}_{\mu}-{\vec p}_{\pi})}.
\end{eqnarray}
The energy  is modified by the ${\vec v}_{\nu}$ dependent
term and particularly the neutrino energy and momentum are 
combined to the small value, $E({p_{\nu}})-
{\vec v}_{\nu}\cdot{\vec p}_{\nu}={m_{\nu}^2 \over E_{\nu}}$. Hence the muon
energy  can be larger  than that of the energy-momentum conserved system and
leads the large finite-size correction.

To compute the amplitude and probability of this component rigorously, we
introduce a correlation function and write the probability, after the spin summations are made,  in the form 
 \begin{align}
&\int  \frac{d{\vec
 p}_l}{(2\pi)^3} \sum_{s_1,s_2}|T|^2 \nonumber\\
&=   \frac{N_2}{E_\nu}\int d^4x_1 d^4x_2 
e^{-\frac{1}{2\sigma_\nu}\sum_i ({\vec x}_i-\vec{x}_i^{\,0})^2} \Delta_{\pi,l}(\delta x)
e^{i \phi(\delta x)},
\label{probability-correlation} 
\end{align}
where $N_2=g^2
\left({4\pi}/{\sigma_{\nu}}\right)^{\frac{3}{2}}V^{-1}$, $V$ is
a normalization volume for the initial pion, $\vec{x}_i^{\,0} = \vec{\text{X}}_{\nu} + {\vec
v}_\nu(t_i-\text{T}_{\nu})$, $\delta x
=x_1-x_2$, $\phi(\delta x)=p_{\nu}\!\cdot\!\delta x $
  and 
\begin{align}
\Delta_{\pi,l} (\delta x)=
 {\frac{1}{(2\pi)^3}}\int
{d {\vec p}_l \over E({\vec p}_l)}(2p_{\pi}\cdot p_{\nu} p_{\pi}\cdot
 p_l-m_{\pi}^2 p_l \cdot p_{\nu})\nonumber\\
 \times e^{-i(p_{\pi}-p_l)\cdot\delta x }. 
\label{pi-mucorrelation}
\end{align}
 In this expression the momentum is integrated first, which
is possible because  the probability is finite and integration variables 
can be interchanged. 

The waves of charged lepton at the ultra-violet energy
regions cause a light-cone singularity.
  Changing the variable to $q=p_{\pi}-p_l$ and integrating  the four
  dimensional $q$, we are able to extract the light-cone singularity $\delta(\delta x^2)\epsilon(\delta t)$ \cite{Ishikawa-Tobita-prl,Wilson-OPE}
  easily. The integral from the region $ q^0 \leq 0  $ leads 
$\delta(\delta x^2)\epsilon(\delta t)$ and the terms written by Bessel 
functions, while that from $0\leq q^0 \leq E_{\pi}$ leads the rapidly
  oscillating term . The light-cone singularity  is real without
  oscillation and is extended to infinity and the others  either
  oscillate  or decrease rapidly.  So this expression is useful to find 
the finite-size correction of the probability which is unable to obtain
 with   standard calculations of 
plane waves. In the  first region the energy is not
  conserved, and the integral vanishes at $\text{T}=\infty$ in fact.  
   The second one, on the other hand, is that of conserving the energy 
and  determines the quantities  at $\text{T}=\infty$.  
   This  expression of writing the probability with the light-cone 
singularity converges
  and is valid  in the  kinematical region $2 p_{\pi}\!\cdot\!
  p_{\nu}\leq \tilde{m}_{l}^2$, where $\tilde{m}_l^2 = m_\pi^2 - m_l^2$.

Substituting the expression of $\Delta_{\pi,l} (\delta x)$ into
Eq.\,$(\ref{probability-correlation} )$, 
we have the phase factor of the
neutrino wave at the light-cone position ${\vec x}=c{\vec
n}_{\nu}(t-\text{T}_{\nu})+\vec{\text{X}}_{\nu}$ in the form, $\phi=E_{\nu}(t-\text{T}_{\nu})-{\vec p}_{\nu}\!\cdot\!({\vec
x}-\vec{\text{X}}_{\nu})=\frac{m_{\nu}^2}{2E_{\nu}}(t-\text{T}_{\nu})$
of  a slow angular velocity. 
Next   an integration on  the coordinates, $\vec x_1$ and ${\vec x}_2$, and  
$t_1$ and $t_2$ in the  finite $\text{T}=\text{T}_{\nu}-\text{T}_{\pi}$, 
leads  the slowly decreasing    term    $\tilde g(\text{T},\omega)$
 and the normal term  $ G_0$.  $\tilde
g(\text{T},\omega)$ is generated from the light-cone singularity and related term
and at $\text{T}=\infty $, $\tilde g(\text{T},\omega_{\nu})$ vanishes.
The normal term,    $ G_0$, is from the rest.
 Due to the rapid oscillation in $\delta t$, $G_0$  gets
 contribution  from 
 the microscopic  $\delta t$ region and  is constant in T. This 
term does not depend on $\sigma_\nu$ and agrees with  the normal probability 
obtained with the standard method of using plane waves. 
In the region $2 p_{\pi}\!\cdot\! p_{\nu} >
{\tilde{m}_{l}^2}$,  $\Delta_{\pi,l} (\delta x)$ does not have a 
light-cone singularity and the diffraction term  exists  only in 
the kinematical region 
$2 p_{\pi}\!\cdot\! p_{\nu}\leq {\tilde{m}_{l}^2}$.  
This  region depends upon the charged-lepton mass, hence  the diffraction terms 
of all three mass eigenstates converge in the union of the kinematical 
regions of the three masses,  $2 p_{\pi}\!\cdot\! p_{\nu}\leq
{\tilde{m}_{\mu}^2}$. The diffraction terms is  applied in this region
in this paper.

  We compute  the total 
probability next. From the integration of neutrino's coordinates
$\vec{\text{X}}_{\nu}$  the total volume $V$ is obtained and  cancelled with 
the normalization of the initial pion state. The total
probability, then, becomes  sum of the normal term $G_0$ and the
diffraction term $\tilde g(\text{T},\omega_\nu)$, 
\begin{align}
\label{probability-3}
P=N_3\int \frac{d^3 p_{\nu}}{(2\pi)^3}
\frac{p_{\pi}\! \cdot\! p_{\nu}(m_{\pi}^2-2p_{\pi}\! \cdot\! p_{\nu}) }{E_\nu}
 \left[\tilde g(\text{T},\omega_{\nu}) 
 +G_0 \right],
\end{align}
where $N_3 = 8\text{T}g^2 \sigma_\nu$ and $\text{L} = c\text{T}$ is the
length of the decay region. $P$ is the neutrino flux when it interacts
with the physical state of finite size target of $\sigma_{\nu}$. 
  At finite T, the flux  has the
diffraction component, which is caused by the superposition of waves 
and stable under the variation of the
pion's momentum.            
At $\text{T} \rightarrow \infty$, the diffraction term vanishes and the
probability $P$ agrees with  the value of the standard
calculation using plane waves.

 \begin{figure}[t]%
  \centering{
   {\includegraphics[angle=-90,scale=.44]{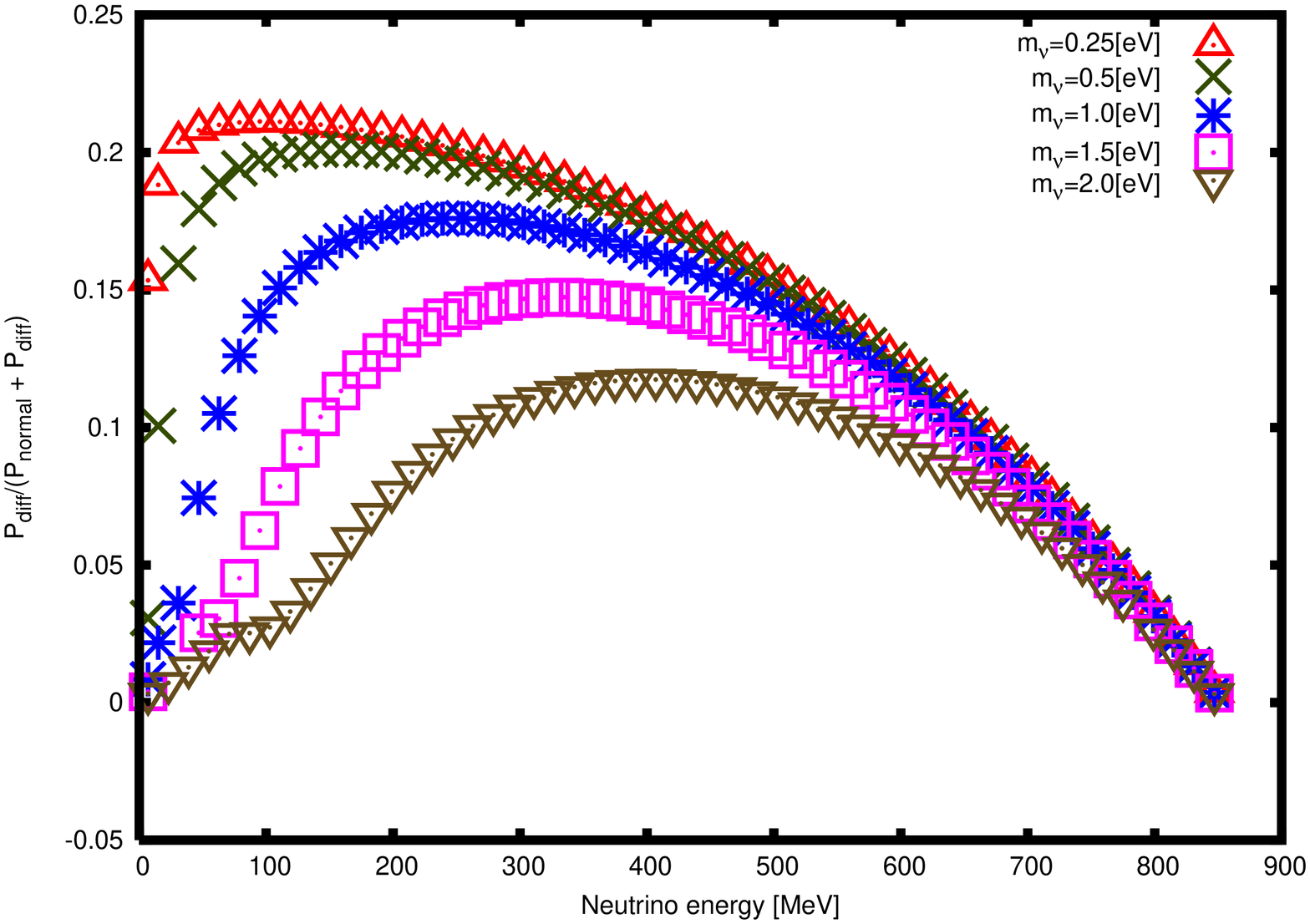}}%
   {\includegraphics[angle=-90,scale=.22]{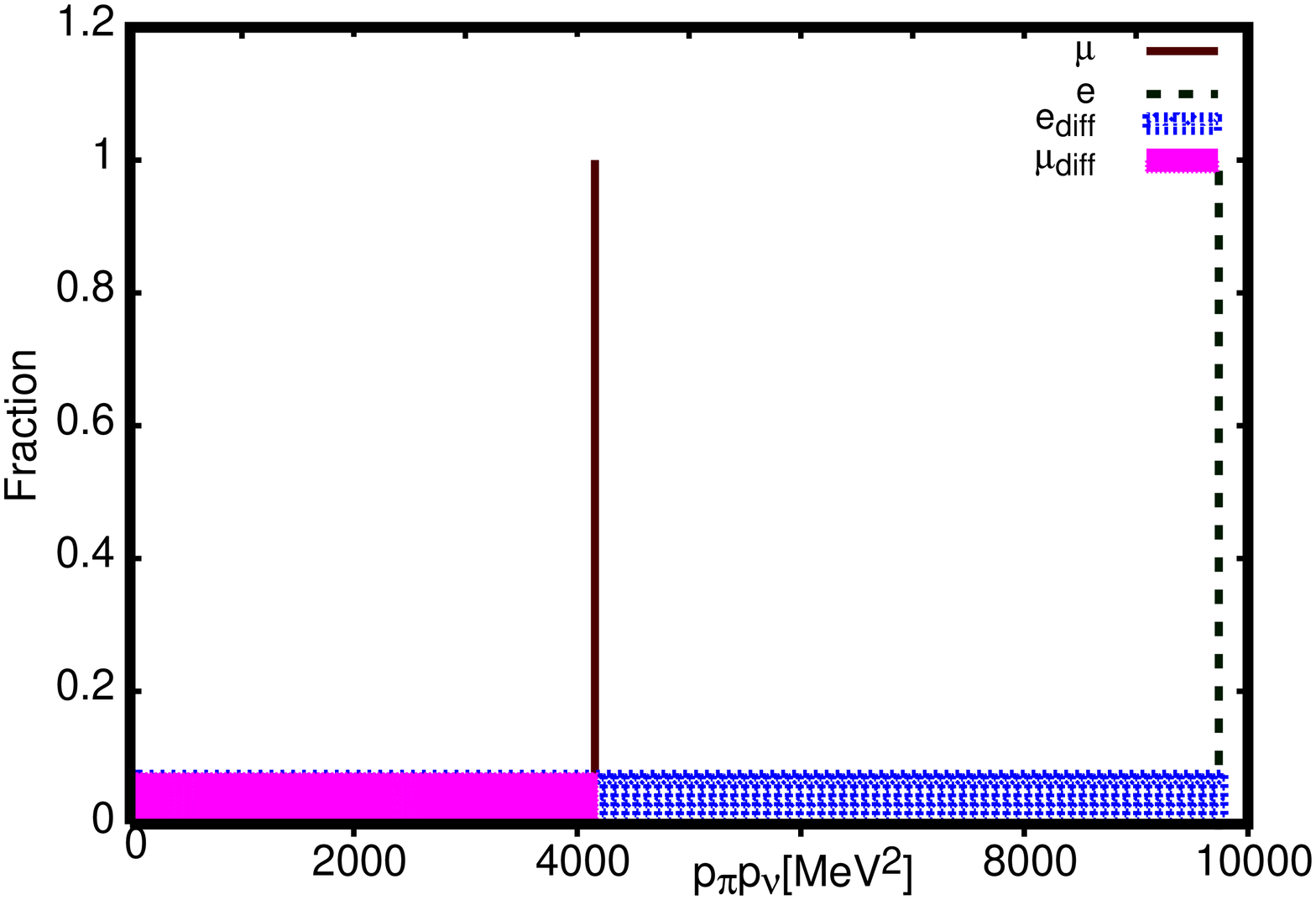}}%
  }
  \caption{The neutrino energy and  $p_{\pi}\cdot p_{\nu}$
  dependences of the fraction of $P_{diff}$ are  given 
in 1-a and 1-b.   The horizontal axis 
  shows $E_\nu$ in~[MeV].   $m_\nu = 0.25-2.0$~[eV/$c^2$], $E_\pi = 2$~[GeV],
  and $\text{L} = 50$~[m]. $p_{\pi}p_{\nu}$ dependence of the
  diffraction term is  broad and that  of the normal term is  narrow.}
  \label{fig:total-int-2}
 \end{figure}%

Now we study  each term in Eq.\,$(\ref{probability-3})$. In the normal 
term, energy and momentum are 
approximately conserved, and  $G_0$ has a sharp peak  at 
$p_{\pi}\!\cdot\! p_{\nu}={\tilde{m}_l^2/2}$. Hence the factor
 $m_{\pi}^2-2p_{\pi}\cdot p_{\nu}$ in Eq.\,$(\ref{probability-3})$ becomes
 $m_l^2$ and the rate is proportional to $m_l^2$. Integration of  
the neutrino's angle leads
  this
  integral  independent of  the angle width, as far as it
  include the narrow peak. The value   is independent also  from 
$\sigma_{\nu}$, which is consistent
 with  the condition for the stationary state \cite{Stodolsky}, and
  agrees with the value of the ordinary method, which  has the
  suppression of   the electron mode. On the 
other hand,   the diffraction component is present in the wide
kinematical region,  
$|\vec{p}_{\nu}|(E_{\pi}-|\vec{p}_{\pi}|)\leq p_{\pi}\!\cdot\! p_{\nu}
\leq {\tilde{m}_l^2/2}$ and depends on $\sigma_{\nu}$.  The size of the nucleus 
of the mass number $A$, 
$\sigma_{\nu}= A^{\frac{2}{3}}/m_{\pi}^2$ 
makes  the value,  $\sigma_{\nu}= 5.2/m_{\pi}^2$ for the
${}^{12}$C nucleus. Since the energy non-conserving states contribute,
the diffraction component has a larger value of  $m_{\pi}^2-2p_{\pi}\!\cdot\! p_{\nu}$ in Eq.$(\ref{probability-3})$ than  $m_l^2$. Hence  the
 branching ratio to  the electron mode  can become much larger than that of 
the normal
 terms and is determined by the
 integral of the diffraction component over the  angle covered by the
 detector in the above region. This value is sensitive to the geometry
 of the experiment  and  could be much larger than  
$10^{-4}$, if the distance between the decay region and the detector is small.  
 
 The energy and $p_{\pi}\!\cdot\! p_{\nu}$ dependences of the fraction of the 
diffraction component  
are 
 presented  in 
Fig.\,\ref{fig:total-int-2} for
$m_{\nu}=0.25-2 \,[\text{eV}/c^2]$, ${ E_\pi = 2}$\,[GeV], and $L=
50$\,[m]. The energy distribution of the diffraction term has a peak at ${ E_\nu =
 100-200}$\,[MeV], and the $p_{\pi}\!\cdot\! p_{\nu}$ distribution is broad in the
 diffraction component and is sharp in the normal component.

From Fig.\,$(\ref{fig:total-int-2})$,  there is an excess
   due to the diffraction in the  short distance region and the
 maximal excess is about $0.2$ of the normal term.
 The diffraction term is slowly varying with both the distance and energy.
The typical length $\text{L}_0$ of this universal behavior  is  
$\text{L}_0~[\text{m}] ={2E_{\nu} \hbar c / m_{\nu}^2 }= 20\times {E_{\nu}[50\text{MeV}]/
 m_{\nu}^2[\text{eV}^2/c^4]}$.
The diffraction is stable with  the energy, hence  a measurement with  a 
finite uncertainty of the neutrino's  energy  $\Delta E_{\nu}$,
 which is of the 
order $0.1 \times E_{\nu}$, does not change the value much.  
For instance, with the uncertainty  $5$\,[MeV] of the energy
$50$\,[MeV]   the diffraction component does not change at all.
For a larger
value of energy uncertainty, too, the diffraction is stable 
under a change of the energy and is computed from 
Eq.\,(\ref{probability-3}).

\section{Electron neutrino in pion decay.} Formulae for general three families  
are easily obtained from  the one species formula.  Since the light-cone
singularity in $\Delta_{\pi,l} (\delta x)$ and the 
diffraction term have the universal form that is independent of the 
mass of the charged lepton, and depends upon the
 absolute neutrino mass, the diffraction term to  the electron neutrino  
 is obtained from  the mixing matrix
 $U_{i,e}$, Eq.\,$(\ref{diffraction})$. Hence the 
flux  of the electron neutrino is written with  the 
average mass $\bar m_\nu^2$, 
$ \sum_i \tilde g(\text{T},\omega_{\nu_i})|U_{i,e }|^2= \tilde g(\text{T},\bar
  \omega_\nu),\bar \omega_\nu={{\bar m_\nu}^2}/{2E_{\nu}}$. The average value 
coincides with $m_{{\nu}_i}$ if the $\delta
m_{\nu}^2$ are much smaller than the average value.

 \begin{figure}[t]%
  \begin{center}
   \subfigure[LSND]{\includegraphics[angle=-90,scale=.3]{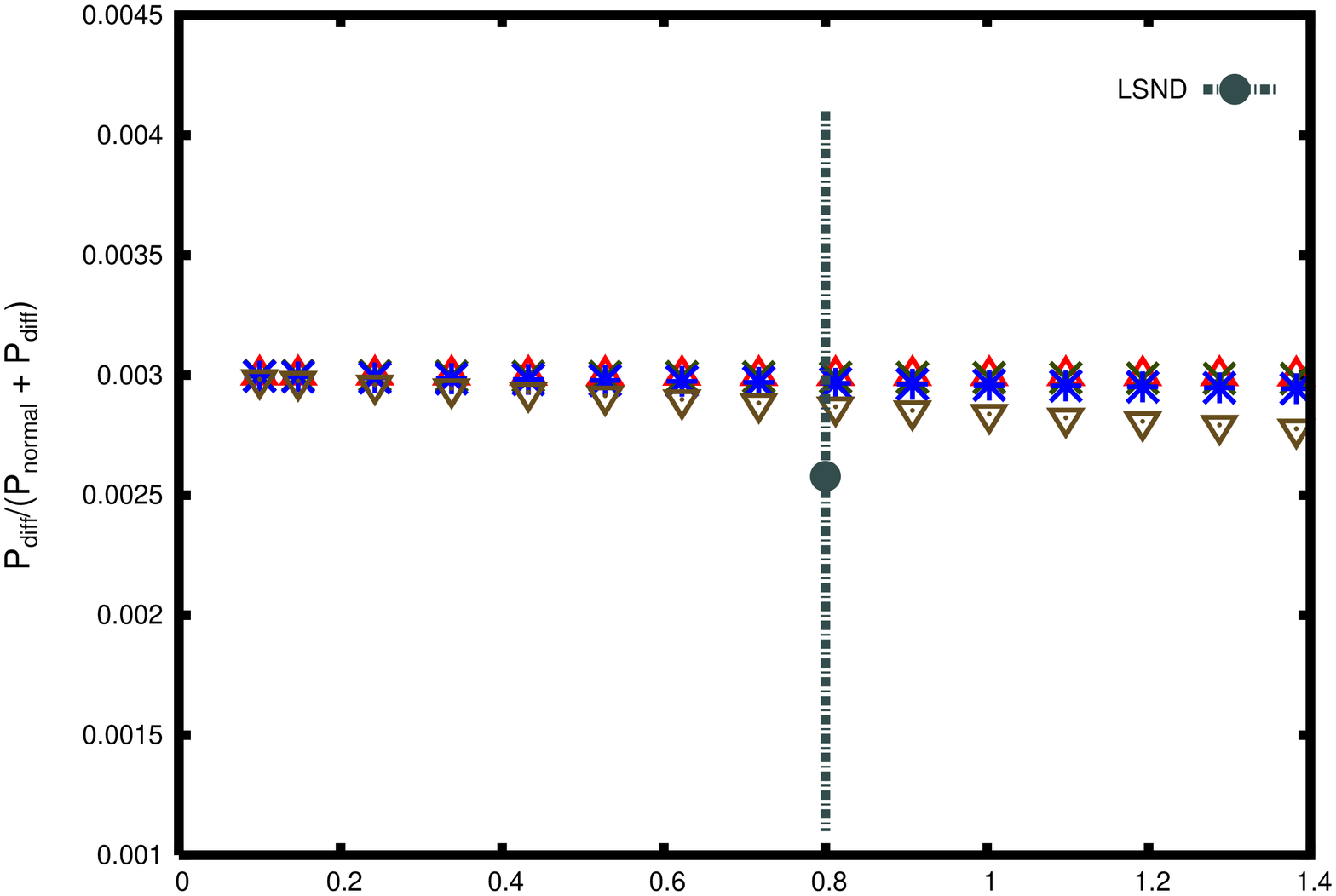}}%
   \subfigure[TWN]{\includegraphics[angle=-90,scale=.3]{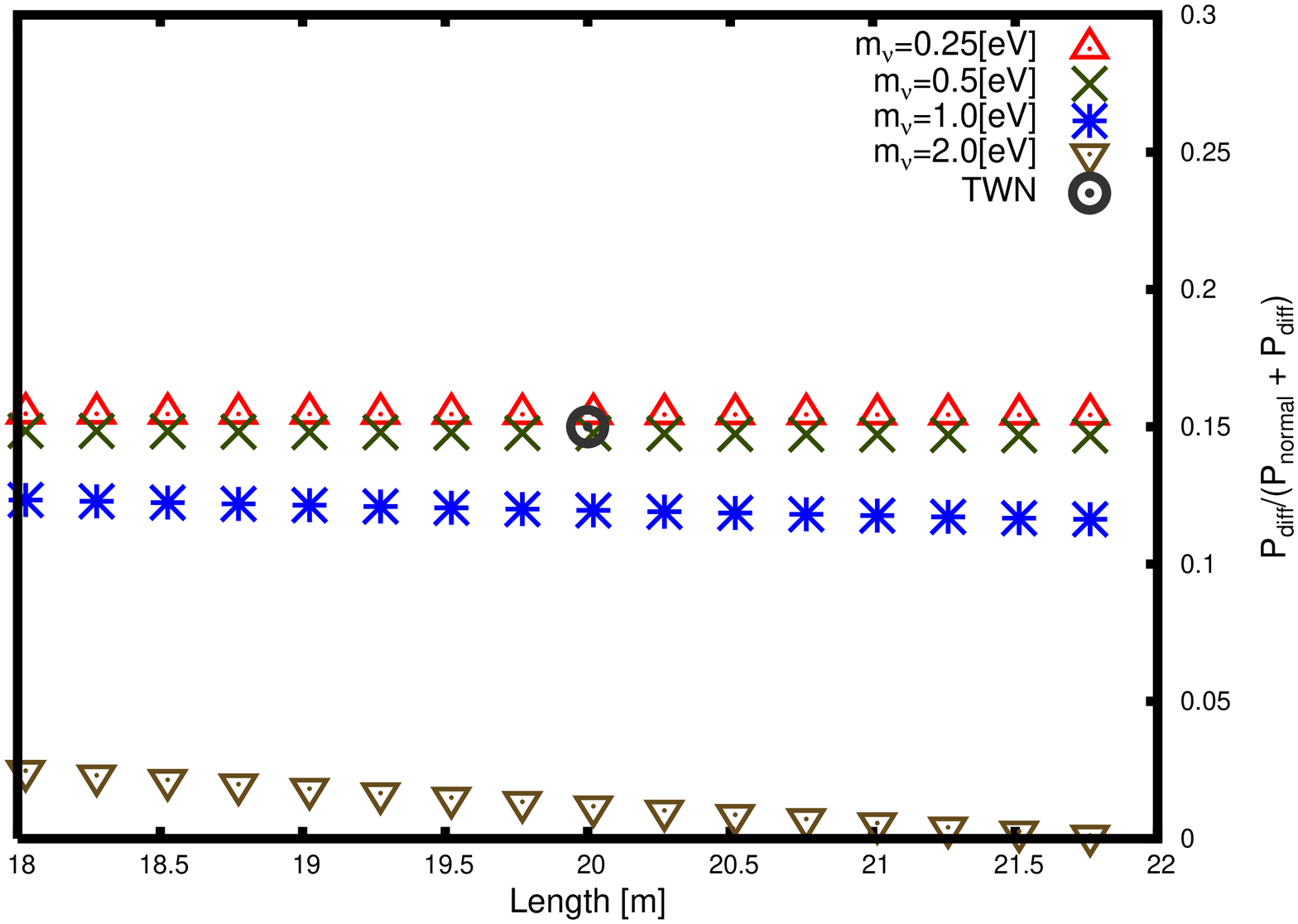}}%

  \end{center}
  \caption{The  fraction  of the diffraction probability  at a finite 
distance L are  
  given for LSND (2-a)and TWN (2-b).   The horizontal axis 
  shows the distance in~[m] and the vertical axis shows fractions. The
  two values   are very different 
 but are consistent and agree with the
  diffraction predictions.
  $\bar{m}_\nu = 0.25-2.0$~[{eV}/$c^2$], $P_\pi = 350$~[MeV/$c$] in LSND
  and $ 300$~[MeV/$c$] in TWN, and 
$E_\nu = 60$~[MeV] in LSND and $ 100$~[MeV] in TWN.}
  \label{fig:total-int-1}
 \end{figure}%

Historically using  neutrinos in pion decays,  
TWN  \cite{excess-two-neutrino} proved  
that  two neutrinos are different, and LSND
\cite{excess-LSND} claimed observation   of
$\nu_{\mu} \rightarrow \nu_e$ transition   with   a large
$\delta m_{\nu}^2=0.1-1.0\,[\text{eV}^2/{c^4}]$. Unfortunately this value  is
inconsistent with MiniBooNE \cite{excess-MiniBoone} which was designed to
to test LSND  and the values $\delta
m_{\nu}^2=7.6\times10^{-5},2.4\times10^{-3}\,[\text{eV}^2/{c^4}] $ from other 
experiments \cite {particle-data}. TWN found also signals of electron neutrino flux at an
order of $10-20$ per cent, which is too large for the flavor oscillation
assumption. So electron neutrino events of LSND and TWN  are
inconsistent with other neutrino
experiments, if they  are attributed to flavor
oscillations. They have been puzzles for quite some time. 

Now we compare  the theoretical value  of electron neutrino 
events  with experiments. The normal component  is
negligibly small due to helicity suppression and the diffraction 
component $P_{diff}$  is  finite and is studied. The neutrino flux  in 
the decay pipe area  is computed from the diffraction formula and the 
neutrino expressed by the wave packet of $\sigma_{\nu}$ determined by the
detector propagates freely in the next area between the decay pipe and 
the detector region.
The distribution is spread with $p_{\pi}\!\cdot\!p_{\nu}$  and has 
different geometrical
behaviors from the normal component in the detector. 
Effects of geometries are important. 
 The geometries of TWN and  LSND are similar and 
 the lengths between the
pion source and the neutrino detector, $L_{d-s}$, and those  
of the decay region, $L_{d-reg}$, 
 are given in Table \ref{geometry-experiments}.
\begin{table}[t]
\caption{Geometries of experiments}
\label{geometry-experiments}
\centering{
\begin{tabular}[t]{ccc}\toprule
 Experiment & $L_{d-s}$ & $L_{d-reg}$\\\midrule
TWN & 30\,[m] & 20\,[m]\\
LSND & 30\,[m] & 0.7\,[m]\\
CDHSW(N) & 120\,[m] & 50\,[m]\\
MiniBooNE & 500\,[m]& 25\,[m](diameter = 12.2\,[m])\\
CDHSW(F) & 600\,[m] & 50\,[m]\\
KARMEN & -  & Stopped pion or muon decay \\
\bottomrule
\end{tabular}
}
\end{table}

\begin{table}[t]
\caption{Experimental values and theoretical values}
\label{theory-experiment}
\centering{
\begin{tabular}[t]{ccc}\toprule
 Experiment & $P_{\nu_e}/P_{\nu_{\mu}}$ (Exp) &$P_{diff}/P_{normal}$ (Th) \\\midrule
TWN & $0.18$ & 0.17\\
LSND &  $(2.6\pm 1.0\pm0.5)\times10^{-3}$& $2.8\times10^{-3}$\\
CDHSW(N) & ?  & 0.2-0.5\\
CDHSW(F) & ? $ $ &   $0.02-0.05$\\
MiniBooNE,  KARMEN & 0 & $\approx 10^{-5}$\\
\bottomrule
\end{tabular}
}
\end{table}
$0.7\,[\text{m}]$ of $L_{d-reg}$ is the length of the decay region in air, while 
the whole decay region is $1.8\,[\text{m}]$. 
Since the length between the detectors and the sources 
of MiniBooNE and CDHSW(F) \cite{neutrino-CDHS} 
is   as large as 10
times those of   TWN  and LSND, 
the contributions of diffraction terms are $1 \over
100$ of TWN  and LSND. At CDHSW(N), the length is $4$ times  of  TWN
and LSND but  the energy is higher and about the same magnitude of the 
diffraction term is obtained.  The diffraction term
comes from the tree diagram and is 
suppressed strongly in  matter, KARMEN \cite{neutrino-KARMEN}. The
diffraction component is important in decay in air but is complicated  in
decay in matter. Its magnitude will be studied in a future publication.
\begin{figure}
\centering{\includegraphics[scale=.35,angle=-90]{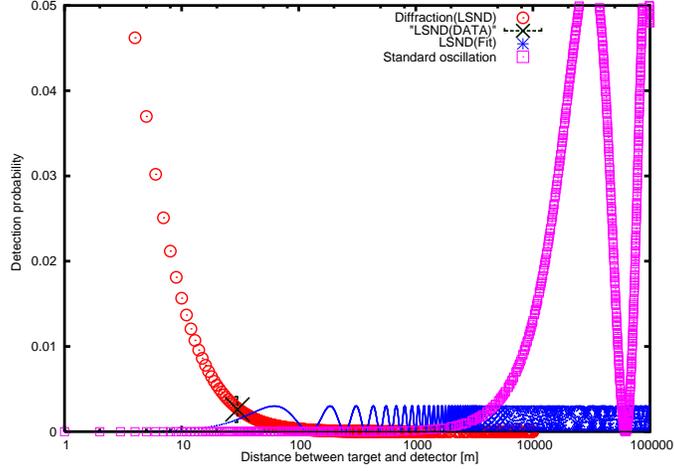}
\caption{Electron neutrino flux of the neutrino diffraction  is compared 
with flavor oscillations. The horizontal axis shows the distance in [m]
 and the vertical axis shows the fraction of the electron neutrino.
 The diffraction agrees with the LSND.  
The value from $\Delta m^2$ of the standard 
oscillation parameters becomes negligible at 30 M. To fit with the
 experiment by the flavor oscillation, a sterile neutrino
 of $\Delta m^2 \approx 1 (eV/c^2)^2$ is necessary. }
}
\end{figure}
We compute  the fraction of the diffraction   components
by taking into account  the geometry. In  LSND,  the decay region in
air is about $0.7\,[\text{m}]$ long and the 
detector is located $30\,[\text{m}]$ away from the decay region.  In this case,  the
ratio becomes  $0.28$ percent at $\bar{m}_{\nu}=1\,[\text{eV}/c^2]$. 
In TWN,  the decay region is about $20\,[\text{m}]$ long and the 
detector is located $10\,[\text{m}]$ away from the decay region. The neutrino detection
angle is much wider and the ratio  becomes  about $17$ percent at 
$\bar{m}_{\nu}=1\,[\text{eV}/c^2]$.  In CDHSW, the energies of pion and neutrino are
higher and iron is used for the detector. Including these effects,
theoretical values are obtained.

The  ratios  $P_{diff}/ P_{normal}$ of our theory and
$P_{\nu_e}/P_{\nu_{\mu}}$ of experiments 
are summarized in Table $\ref{theory-experiment}$. Uncertainties  in
TWN are 
not known and ignored.
The fractions of the diffraction components obtained in this way are 
 presented  in 
Fig.\,\ref{fig:total-int-1} for the mass of the neutrino,
 $\bar{m}_{\nu}=0.25-2\,[\text{eV}/c^2]$, and the pion momentum ${ p_\pi =
 350}$\,
[MeV/c] in LSND and ${300}$\,
[MeV/c] in TWN, and the neutrino
 energy $E_\nu = 60$\,[MeV] in LSND and $=100$\,[MeV] in TWN. 
 The fractions of the theory  agree with those of the experiments using
 the same $\bar m_{\nu}$.

The magnitude of electron neutrino
flux  agreed   with   LSND in flight  and TWN 
but not to others including MiniBooNE. Since matter effect in LSND at
rest is complicated, we study only LSND in flight here.
In Fig.3, the electron neutrino fraction from the neutrino diffraction is 
compared with  those of the flavor oscillations of standard parameters 
of $\Delta m^2$ and of a
sterile neutrino.  The neutrino diffraction explains naturally the
excess of the electron neutrino. Thus the puzzles are resolved 
and consistent understanding  of  the neutrino experiments are
obtained.
  Furthermore,  information  on the absolute neutrino mass is
  obtained \cite {particle-data,Tritium,WMAP-neutrino}. 

\section{Summary and implications.} 
It is found that the neutrino detection rate has the unusual finite-size 
correction in the macroscopic distance. Instead of the standard S-matrix
$S[\infty]$, which is useless to study  the finite-size correction, 
we studied  the S-matrix of the finite time interval $S[\text{T}]$ and 
computed the finite-size correction of the detection probabilities which
are measured in experiments.   
The asymptotic boundary
conditions must be satisfied for the scattering or decay processes,
hence the wave packets are necessary  for  $S[\text{T}]$ and are used.  
One feature is that S[\text T] does not commute with the free 
Hamiltonian $H_0$ and  both  final states of
conserving the energy and those of non-conserving energy contribute.
The rate  from the former is independent of both  $\text T$ and
$\sigma_{\nu}$ and agrees   with that of the
standard method, while that from the latter depends on $\text T$ and
$\sigma_{\nu}$ and does not satisfy various relations of the former such
as the helicity suppression, symmetry between the neutrinos and charged
leptons,  and others.

The electron neutrino flux  due to  the diffraction  was computed 
in the geometries of   LSND and TWN and  excellent agreements between
the theory and  experiments
are obtained. Because 
the electron neutrino flux due to diffraction is determined by the average
neutrino mass $\bar m_\nu$, the excesses of the electron neutrino 
  suggest that $\bar m_\nu$ is around $0.25-1.0\,[\text{eV}/c^2]$. 
If the $\bar m_\nu$ is in this range and $\delta m^2_\nu$ of the experiments 
are used,  LSND and TWN electron events are attributed to the diffraction
component but not  to the flavor oscillation. Thus  the controversy 
related with  LSND and TWN are resolved. 
 Some information on the absolute neutrino mass 
is obtained, furthermore.

The confirmation of the neutrino diffraction would be made in a new 
experiment from the unique  behaviors  of the neutrino spectrum  on the $p_{\pi}\!\cdot\!p_{\nu}$, energy, and angle. 
  Since the helicity suppression does
not work in the diffraction component, the flux  of the electron
neutrino becomes about $10-20$ \%\ of that of the muon neutrino at the maximum
value. Furthermore it has the unique properties that are very different
from the normal term.
If clear distinctions are  observed on the  dependences of these variables
and  agreements  with the theory are   found, the neutrino
diffraction will be confirmed.
Since at much larger
distance than the above length,  the diffraction component 
disappears and only the normal   flavor oscillation terms remain,
the experiment to measure the excess due to 
the diffraction component
 should be made at a short distance.

The present results may exhibit  violations  of two important relations,
the unitarity and lepton number conservation. However that is not the
case and they are not violated.  
First, from the behavior of the diffraction
component that  the detection rate is decreasing with $\text T$, the 
neutrino number  looks decreasing and an unitarity  is violated
superficially.  Since the neutrino  is detected  at the  distant position, the
detection time is delayed. The  retarded and interference effects made  
the detection probability deviate from the neutrino number.  Hence the 
decreasing behavior of  $P({\text T})$  is in-connected with both the
decreasing neutrino number and unitarity. In the text,   
$P({\text T})$ is computed with  $S[\text{T}]$ that satisfies the proper
boundary condition of the experiments and  the unitarity condition  
$S[\text{T}]^{\dagger}S[\text{T}]=1$. If the
$\text T$ dependent neutrino probability were computed with $S[\infty]$,
this would  not  have decreased with $\text T$. However, the 
$S[{\infty}]$  does not satisfy the proper boundary condition for the 
neutrino at $\text T$ and is neither applied nor  compared with the 
probability measured by the experiment. $S[\text {T}]$, on the other
hand, satisfies the proper boundary condition and   is  used for
comparisons with the experiments at the finite $\text T$.  One of the
important findings of the present work is that $S[\text {T}]$ gives the 
different results from
$S[\infty]$ even in the macroscopic $\text T$. 
Second, asymmetric behaviors of $\tilde g(\text{T},\omega)$ lead that 
the neutrino detection rates are very different from those of the  charged
leptons.  This is not  in-consistent  
with the lepton number conservations but means that the detection
probability is different from the neutrino   number owing to the
finite-size correction caused by 
retarded and interference effects. When either the neutrinos or the
charged leptons are observed, their detection rates  are different. 
Nevertheless, when  both of neutrino and charged lepton are observed 
simultaneously, they have the same rates, due to the lepton number 
conservations.

The shorter life time of the pion in the situation when  the neutrino is
observed at a small distance than that of the free pion has not been
tested and will be confirmed in a future experiment. Then if the muon is
measured simultaneously, this muon  shows the same behavior from the 
lepton number conservation. The detection rate of the muon depends on the
boundary condition on the neutrino. 

In this paper we ignored the higher order effects such as the pion life 
time and the pion mean free path in studying the quantum effects. We will
study these problems, the muon decay, and other large scale physical 
phenomena of low energy neutrinos in subsequent papers.

\section*{Acknowledgements.} Authors  thank
Drs. Kobayashi, Maruyama, Nakaya, Nishikawa for useful discussions on 
the near detector of T2K experiment, Drs. Asai, Komamiya, Kobayashi, Minowa, 
Mori, Yamada for useful discussions on interferences, and Dr. Eun-Kyung
Park for careful reading of the manuscript. 
{}

\end{document}